\documentclass[sigconf,natbib=true,screen=true,nonacm]{acmart}

\copyrightyear{2025}
\acmYear{2025}
\setcopyright{cc}
\setcctype{by}
\acmConference[SIGIR '25]{Proceedings of the 48th International ACM SIGIR Conference on Research and Development in Information Retrieval}{July 13--18, 2025}{Padua, Italy}
\acmBooktitle{Proceedings of the 48th International ACM SIGIR Conference on Research and Development in Information Retrieval (SIGIR '25), July 13--18, 2025, Padua, Italy}\acmDOI{10.1145/3726302.3729915}
\acmISBN{979-8-4007-1592-1/2025/07}
\usepackage{acronym}
\usepackage[skip=0pt]{caption}
\usepackage[inline]{enumitem}
\usepackage{enumitem}
\usepackage{booktabs} 
\usepackage{multirow}
\usepackage{array}
\usepackage{pifont}
\usepackage{xspace}
\usepackage{subcaption}
\usepackage{makecell}
\usepackage{CJKutf8}
\usepackage{tcolorbox}
\usepackage[para]{footmisc}
\usepackage{amsfonts}
\usepackage{rotating}

\acrodef{QF-DC}{query filtering mechanism based on document relevance and conversation alignment}

\acrodef{RAG}{retrieval-augmented generation}

\acrodef{RCD}{Retrieval from Conversational Dialogues}
\acrodef{PSC}{proactive search in conversations}
\acrodef{CPS}{conversational proactive search}
\acrodef{PS}{proactive search}

\acrodef{GR}{generative retrieval}
\acrodef{LLM}{large language model}
\acrodef{QPP}{query performance prediction}
\acrodef{IR}{information retrieval}
\acrodef{NLP}{natural language processing}
\acrodef{PEFT}{parameter-efficient fine-tuning}
\acrodef{ICL}{in-context learning}
\acrodef{LoRA}{low-rank adaptation}

\acrodef{RR}{reciprocal rank}
\acrodef{AP}{Average Precision}
\acrodef{nDCG}{normalized discounted cumulative gain}

\acrodef{HSD}{Tukey's honestly significant difference}
\acrodef{ANOVA}{analysis of variance}

\acrodef{UEF}{utility estimation framework}
\acrodef{QPP-PRP}{pairwise rank preference-based QPP}
\acrodef{M-QPPF}{multi-task query performance prediction framework}
\acrodef{WRIG}{weighted relative information gain-based model}
\acrodef{NLP}{natural language processing}
\acrodef{CIS}{conversational information seeking}
\acrodef{ANCE}{Approximate nearest neighbor Negative Contrastive Estimation}

\acrodef{RL}{reinforcement learning}

\newcommand\setItemnumber[1]{\setcounter{enumi}{\numexpr#1-1\relax}}

\newcommand{\header}[1]{\vspace*{1mm}\noindent\textbf{#1}.}

\ccsdesc[500]{Information systems~Evaluation of retrieval results}

\keywords{Re-ranking, Relevance prediction, Relevance judgment}

\author{Chuan Meng}
\orcid{0000-0002-1434-7596}
\affiliation{%
  \institution{The University of Edinburgh}
  \city{Edinburgh}
  \country{United Kingdom}
}
\email{chuan.meng@ed.ac.uk}

\author{Jiqun Liu}
\orcid{0000-0003-3643-2182}
\affiliation{
  \institution{University of Oklahoma}
  \city{Norman}
  \country{United States}
}
\email{jiqunliu@ou.edu}

\author{Mohammad Aliannejadi}
\orcid{0000-0002-9447-4172}
\affiliation{
  \institution{University of Amsterdam}
  \city{Amsterdam}
  \country{The Netherlands}
}
\email{m.aliannejadi@uva.nl}

\author{Fengran Mo}
\orcid{0000-0002-0838-6994}
\affiliation{%
  \institution{Université de Montréal}
  \city{Montréal}
  \country{Canada}
}
\email{fengran.mo@umontreal.ca}

\author{Jeff Dalton}
\orcid{0000-0003-2422-8651}
\affiliation{%
  \institution{The University of Edinburgh}
  \city{Edinburgh}
  \country{United Kingdom}
}
\email{jeff.dalton@ed.ac.uk}

\author{Maarten de Rijke}
\orcid{0000-0002-1086-0202}
\affiliation{%
  \institution{University of Amsterdam}
  \city{Amsterdam}
  \country{The Netherlands}
}
\email{m.derijke@uva.nl}

\begin{document}

\title[Re-Rankers as Relevance Judges]{Re-Rankers as Relevance Judges}
\renewcommand{\shortauthors}{Chuan Meng et al.}

\begin{abstract}
Using large language models (LLMs) to predict relevance judgments has shown promising results.
Most studies treat this task as a distinct research line, e.g., focusing on prompt design for predicting relevance labels given a query and passage. 
However, predicting relevance judgments is essentially a form of \textit{relevance prediction}, a problem extensively studied in tasks such as re-ranking.
Despite this potential overlap, little research has explored reusing or adapting established re-ranking methods to predict relevance judgments, leading to potential resource waste and redundant development.
To bridge this gap, we reproduce re-rankers in a \textit{re-ranker-as-relevance-judge} setup.
We design two adaptation strategies: (i) using binary tokens (e.g., ``true'' and ``false'') generated by a re-ranker as direct judgments, and (ii) converting continuous re-ranking scores into binary labels via thresholding. 
We perform extensive experiments on TREC-DL 2019--2023 with 8 re-rankers from 3 families, ranging from 220M to 32B, and analyse the evaluation bias exhibited by re-ranker-based judges.
Results show that re-ranker-based relevance judges, under both strategies, can outperform UMBRELA, a state-of-the-art LLM-based relevance judge, in around 40\%--50\% of the cases; they also exhibit strong self-preference towards their own and same-family re-rankers, as well as cross-family bias.
\end{abstract}


%

\maketitle
\acresetall

\section{Introduction}
Relevance judgments, which map each query to the documents that should be retrieved for it~\cite{soboroff2025don}, play a critical role in information retrieval (IR). 
Accurate relevance judgments are essential for both training and evaluating ranking systems~\cite{schnabel2025multi}.
However, the manual annotation of relevance judgments is labour-intensive~\cite{alaofi2024llms}.
Recently, the IR community has witnessed a surge in the use of large language models (LLMs) for automatically predicting relevance judgments~\cite{dewan2025true,merlo2025cost,sakai2025opensource,arabzadeh2025benchmarking,farzi2025does,upadhyay2024umbrela,faggioli2023perspectives,thomas2024large}, which has shown promising results~\cite{thomas2024large}.
%

Most studies tend to treat LLM-based relevance judging as a distinct line of research~\cite{faggioli2023perspectives}. 
However, \citet{soboroff2025don} argues that ``asking a computer to decide if a document is relevant is no different from using a computer to retrieve documents and rank them in order of predicted degree of relevance,'' and that both are ``essentially predictions of relevance.''
Similarly, \citet{clarke2024llm} argue that ``there is no fundamental difference between an LLM-based relevance assessment and an LLM-based re-ranking method.''

\header{Motivation}
This work is driven by two key motivations:
\begin{enumerate*}[label=(\roman*)]
\item Despite the conceptual overlap between LLM-based relevance judgment prediction and ranking, little research has investigated how established ranking methods, such as re-rankers~\cite{ma2024fine,nogueira2020document}, can be adapted for predicting relevance judgments. 
Building LLM-based judges while neglecting these ranking methods may lead to duplicated efforts and the underutilisation of mature relevance prediction resources. 
\item Although many studies~\cite{dietz2025llm,soboroff2025don,dietz2025principles,clarke2024llm,faggioli2023perspectives} warn that using the same LLM for both ranking and evaluation may cause ``circularity'' (i.e., overinflated ranking performance), there is a lack of empirical studies looking into this issue.
\end{enumerate*}

%


%
%
%

\header{Research goal}
In this paper, we examine \textit{to what extent established findings on re-ranking are generalisable to the ``re-ranker-as-relevance-judge'' setup}.
%
In other words, this work reproduces existing re-ranking methods under a new setting, namely as ranking evaluators.
We revisit the following findings from the literature on re-ranking:
\begin{enumerate*}[label=(\roman*)]
    \item re-rankers based on larger language models generally outperform those based on smaller models~\citep{weller2025rank,ma2024fine,nogueira2020document};
    \item reasoning-based re-rankers, which generate explicit reasoning chains before making relevance predictions, generally perform better than non-reasoning-based ones~\citep{weller2025rank,yang2025rank}; and
    \item reasoning-based re-rankers demonstrate robustness~\citep{yang2025rank}. 
\end{enumerate*} 

\header{Challenge and solution}
The main challenge lies in the output format mismatch between relevance judgment prediction and re-ranking: the former assigns discrete labels (e.g., relevant or irrelevant) to query--document pairs, whereas the latter typically outputs continuous relevance scores to generate ranked lists.
To bridge this gap, we devise two adaptation strategies:
\begin{enumerate*}[label=(\roman*)]
\item \textit{Direct generation}: for re-rankers that compute relevance scores from the output logits of special tokens (e.g., ``true'' and ``false'')~\cite{weller2025rank,nogueira2020document}, we directly use the final generated token as the predicted relevance judgment; and
\item \textit{Score thresholding}: converting continuous re-ranking scores into binary labels via a threshold.
\end{enumerate*}

\header{Scope}
Based on the two adaptation strategies described above, our study investigates the effectiveness of re-ranker-based relevance judges under each strategy.
We first examine the following research questions:
\textbf{RQ1:} Do re-rankers generalise as effective relevance judges through direct generation?
\textbf{RQ2:} How do re-rankers perform when using score thresholding?
Beyond effectiveness, using re-rankers as relevance judges may introduce evaluation bias and circularity~\cite{dietz2025llm,soboroff2025don,dietz2025principles,clarke2024llm,faggioli2023perspectives}.
Therefore, it is important to investigate the risk of bias introduced by re-ranker-based judges, and we further examine the following research question:
\textbf{RQ3:} To what extent do re-ranker-based judges exhibit bias when evaluating re-rankers?

%
%

\header{Experiments}
We conduct experiments on the TREC 2019–2023 Deep Learning (TREC-DL) tracks~\cite{craswell2019,craswell2020,2021craswell,craswell2022,craswell2023}, which provide dense and high-quality human-annotated relevance judgments.
Our study considers 3 representative re-ranker families: monoT5~\cite{nogueira2020document}, Rank\-LLaMA~\cite{ma2024fine}, both widely used in the ranking literature~\cite{weller2025rank,yang2025rank,meng2024ranked}, and Rank1~\cite{weller2025rank}, a recently proposed state-of-the-art reasoning-based re-ranker. 
In total, we include 8 re-rankers across the 3 families, with model sizes ranging from 220M to 32B parameters.
We evaluate the resulting re-ranker-based relevance judges by measuring their agreement with TREC assessors using Cohen’s~$\kappa$ and by comparing system ranking.
To provide a reference point for re-ranker-based relevance judges, we use UMBRELA~\cite{upadhyay2025large,upadhyay2024umbrela}, a state-of-the-art LLM-based relevance judge. 

\header{Lessons}
For RQ1 and RQ2, we find that most findings on re-ranking carry over to the ``re-ranker-as-relevance-judge'' setup under both adaptation strategies.
We find that re-ranker-based relevance judges can achieve better performance than UMBRELA~\cite{upadhyay2024umbrela,upadhyay2025large}, in around 40\%--50\% of the cases.
Reasoning-based re-rankers (Rank1) generally outperform  non-reasoning ones, and exhibit greater robustness across datasets.
Scaling up model size tends to benefit system-ranking quality, but the gains are not always consistent.
Score thresholding is particularly effective for re-rankers whose relevance scores are highly concentrated near the extremes; Rank1 shows substantially stronger robustness to threshold variation than other re-rankers.
For RQ3, we find that re-ranker-based judges exhibit strong biases: they rate their own re-ranker highest, followed by same-family re-rankers and then others; they show cross-family bias (e.g., non-reasoning-based re-rankers prefer other non-reasoning-based families over the reasoning-based Rank1 family) and often overestimate BM25’s retrieval quality compared with UMBRELA.

Overall, the competitive performance of re-ranker-based judges provides empirical evidence suggesting that relevance judging is not a fundamentally new task but rather a specific case of \textit{relevance prediction} (consistent with~\cite{soboroff2025don,clarke2024llm}), indicating that the community can benefit from building relevance judges \textit{on top of} well-established relevance predictors to reduce duplicated effort.
Contemporaneous work~\citep{gienapp2025topic} adapts monoT5 as a judge by adding trainable parameters.



\header{Contributions}
Our main contributions are as follows:
\begin{itemize}[leftmargin=*,nosep]
\item To the best of our knowledge, we are the first to systematically reproduce re-rankers as relevance judgment predictors.
\item We design two adaptation strategies, \textit{direct generation} and \textit{score thresholding}, to enable re-rankers to function as relevance judges.
\item Experimental results show that re-ranker-based relevance judges can achieve better performance than UMBRELA in around 40\%--50\% of the cases while tending to exhibit strong self-preference towards their own and same-family re-rankers, as well as cross-family bias.
We open source supplementary materials at \url{https://github.com/ChuanMeng/reranker-as-judge}.
\end{itemize}

\section{Task Definition: From Re-ranking to Relevance Judging}
Given that pointwise re-rankers are well studied in the IR community~\citep{weller2025rank,ma2024fine,nogueira2020document}, and that many widely used IR evaluation metrics (e.g., Precision, Recall, MRR, and MAP) are commonly applied under binary relevance judgments, this paper focuses on pointwise re-rankers and binary relevance judgments. 

Given a query $q$ and a document $d$, a re-ranker $f$ assigns a continuous relevance score $s = f(q,d) \in \mathbb{R}$, which is used to produce a final ranking by sorting documents according to these scores. 
To adapt a re-ranker $f$ into a relevance judge, $f$ instead outputs a discrete relevance label $l = f(q,d) \in \{0,1\}$, where $l=1$ denotes that $d$ is relevant to $q$, and $l=0$ otherwise.

\section{Reproducibility Methodology}
\label{model}


\subsection{Research questions and experimental design}
\begin{enumerate}[label=\textbf{RQ\arabic*},leftmargin=*]
    \setItemnumber{1}
    \item Do re-rankers generalise as effective relevance judges through direct generation? \label{RQ1}
\end{enumerate}
To address \ref{RQ1}, we use our proposed direct generation strategy (see Section~\ref{sec-adaptation} below for details) to adapt re-rankers as relevance judges and evaluate them in terms of judgment agreement with humans and system ranking. \label{E1}
\begin{enumerate}[label=\textbf{RQ\arabic*},leftmargin=*]
    \setItemnumber{2}
    \item Do re-rankers generalise as effective relevance judges through score thresholding? \label{RQ2}
\end{enumerate}
To address \ref{RQ2}, we use the same setup as \ref{RQ1} but apply the score thresholding strategy~(see Section~\ref{sec-adaptation}) instead.
\begin{enumerate}[label=\textbf{RQ\arabic*},leftmargin=*]
    \setItemnumber{3}
    \item To what extent does a re-ranker-based judge show bias towards evaluating re-rankers? \label{RQ3}
\end{enumerate}
To address \ref{RQ3}, each re-ranker-based judge evaluates its underlying re-ranker and other re-rankers, inducing a ranking (based on an IR evaluation metric) over all re-rankers; we then analyse the bias exhibited by each judge~(see Section~\ref{sec-bias-method}).


\subsection{Experimental setup}
\subsubsection{Re-rankers with varying sizes.}
\label{sec-re-rankers}
We use 3 re-ranker families: monoT5~\cite{nogueira2020document}, RankLLaMA~\cite{ma2024fine} and Rank1~\cite{weller2025rank}.
monoT5 and RankLLaMA are widely-used in the literature~\cite{weller2025rank,yang2025rank,meng2024ranked}, while Rank1~\cite{weller2025rank} is a recently proposed state-of-the-art reasoning-based re-ranker.
All are pointwise re-rankers: given a query and a candidate document, they independently assign a relevance score, and the final ranking is produced by sorting documents according to these scores.
All of them are trained on the training set of MS MARCO V1~\cite{bajaj2018ms}.
\begin{itemize}[leftmargin=*,nosep]
\item \textbf{monoT5}~\cite{nogueira2020document} fine-tunes T5~\cite{raffel2020exploring} to output one of two special tokens, ``true'' or ``false'', depending on whether the document is relevant to the query.
monoT5 applies a softmax over the logits of the ``true'' and ``false'' tokens, and uses the probability assigned to the ``true'' token as the relevance score.
\item \textbf{RankLLaMA}~\cite{ma2024fine} fine-tunes Llama 2~\cite{touvron2023llama2} using LoRA~\cite{hu2021lora} to directly project the representation of the end-of-sequence token to a relevance score.
Unlike monoT5, the relevance scores produced by RankLLaMA are unbounded real-valued scores rather than probabilities.
\item \textbf{Rank1}~\cite{weller2025rank} fine-tunes an LLM to generate reasoning traces before producing one of two special tokens, following the output format ``<think> ... </think> true/false''.
Reasoning traces used for training are generated by DeepSeek’s R1 on MS MARCO.
Similar to monoT5, Rank1 uses the probability assigned to the ``true'' token as the relevance score.
\end{itemize} 

\noindent We also investigate the impact of re-ranker scaling on relevance judgment prediction.
We include \textbf{monoT5-base} (220M), \textbf{monoT5-large} (770M), and \textbf{monoT5-3B};
\textbf{Rank\-LLaMA-7B} and \textbf{Rank\-LLa\-MA-13B};
and \textbf{Rank1-7B}, \textbf{Rank1-14B}, and \textbf{Rank1-32B},
covering 8 re-rankers with varying model architectures and sizes.

\subsubsection{Adaptation methods.}
\label{sec-adaptation}

We devise two strategies for adapting re-rankers as re-ranker-based relevance judges:
\begin{enumerate}[leftmargin=*,label=(\arabic*)]
\item \textit{Direct generation} is specific to monoT5~\cite{nogueira2020document} and Rank1~\cite{weller2025rank}. 
Instead of using the numerical relevance score, we directly map the model's generated special token to a relevance label: if monoT5/Rank1 generates the token ``true'', the document is labelled as ``relevant''; if it generates ``false'', it is labelled as irrelevant.
\item \textit{Score thresholding} applies to all re-rankers. We convert their continuous relevance scores into binary labels by applying a pre-defined threshold: a document is predicted as relevant if its score is greater than or equal to the threshold, and irrelevant otherwise.
\end{enumerate}

\noindent%
We focus on generating binary relevance judgments in this work.
Many widely used IR evaluation metrics, such as Precision, Recall, MRR, and MAP, operate under binary relevance judgments, meaning that this setting covers most common IR evaluation scenarios.
Extending these adaptation strategies to support graded relevance judgments is an interesting direction for future work, but is beyond the scope of this paper.

%
\subsubsection{Bias analysis: Re-ranker-based judges evaluating re-rankers.}
\label{sec-bias-method}

Given the 8 re-rankers described in Section~\ref{sec-re-rankers}, we perform a bias analysis by examining how each re-ranker-based judge evaluates different ranking systems.
For each judge, we curate 9 system runs: a BM25 retriever~\citep{robertson1995okapi}, and BM25 followed by each of the 8 re-rankers.
Using relevance judgments produced by the judge, we compute standard IR evaluation metric values for each run and induce a ranking over the 9 systems.
We then analyse the rankings to characterise the evaluation bias exhibited by each re-ranker-based judge.
See Section~\ref{sec:details} for technical details.

\begin{table}[t!]
\centering
\caption{Statistics of the TREC 2019--2023 Deep Learning (TREC-DL) tracks~\cite{craswell2019,craswell2020,2021craswell,craswell2022,craswell2023}. 
For TREC-DL 2022 and 2023, we follow \cite{upadhyay2024umbrela} and use relevance judgments after removing automatically propagated passages.}
\label{tab:data}
\setlength{\tabcolsep}{3.5pt}
\begin{tabular}{l c c l}
\toprule
Track & \#Runs & \# Queries & Relevance labels (0/1/2/3) \\
\midrule
TREC-DL 2019 & 37 & 43 & \phantom{0}5,158 / 1,601 / 1,804 / \phantom{0,}697 \\
TREC-DL 2020 & 59 & 54 & \phantom{0}7,780 / 1,940 / 1,020 / \phantom{0,}646 \\
TREC-DL 2021 & 63 & 53 & \phantom{0}4,338 / 3,063 / 2,341 / 1,086 \\
TREC-DL 2022 & 60 & 76 & 12,892 / 6,192 / 3,053 / 1,385 \\
TREC-DL 2023 & 35 & 82 & 11,618 / 3,774 / 1,942 / 1,544 \\
\bottomrule
\end{tabular}
\end{table}

\begin{table*}[th!]
\centering
\caption{
Relevance judgment agreement (Cohen’s $\kappa$) between TREC assessors and each re-ranker-based relevance judge adapted via \textit{direct generation}.
The results of UMBRELA, state-of-the-art relevance judge, are included for reference.
The best value in each column is \textbf{boldfaced}, while the second best is \underline{underlined}.
}
\label{tab:dg_cohen}
\begin{tabular}{l ccccc}
\toprule
\multirow{1}{*}{Method} & \multicolumn{1}{c}{TREC-DL 19} & \multicolumn{1}{c}{TREC-DL 20} & \multicolumn{1}{c}{TREC-DL 21} & \multicolumn{1}{c}{TREC-DL 22} & \multicolumn{1}{c}{TREC-DL 23} \\
\midrule 
UMBRELA & \textbf{0.499} & \textbf{0.450} & \textbf{0.492} & \underline{0.421} & \textbf{0.418}  \\
\midrule
monoT5 base & 0.389 & 0.322 & 0.343 & 0.239 & 0.269 \\
monoT5 large & 0.425 & 0.323 & 0.340 & 0.246 & 0.291 \\
monoT5 3B & 0.386 & 0.385 & 0.402 & 0.279 & 0.335 \\
Rank1-7B  & 0.429 & \textbf{0.450} & \underline{0.476} & \textbf{0.425} & 0.382 \\
Rank1-14B  & \underline{0.459} & \underline{0.433} & 0.469 & \underline{0.421} & \underline{0.394} \\
Rank1-32B & 0.452 & 0.421 & 0.455 & 0.420 & 0.380 \\
\bottomrule
\end{tabular}
\end{table*}

\begin{table*}[th!]
\centering
\caption{
Kendall’s $\tau$ correlation coefficients between the system orderings induced by relevance judgments from TREC assessors and those predicted by each re-ranker-based relevance judge adapted via \textit{direct generation}. 
Results are shown for system orderings based on MAP@100 and MRR@10.
The results of UMBRELA, state-of-the-art relevance judge, are included for reference.
The best value in each column is \textbf{boldfaced}, while the second best is \underline{underlined}.
}
\label{tab:dg_kendall}
\setlength{\tabcolsep}{3.7pt} 
\begin{tabular}{l ccccccccccc}
\toprule
\multirow{2}{*}{Method} 
& \multicolumn{2}{c}{TREC-DL 19} 
& \multicolumn{2}{c}{TREC-DL 20} 
& \multicolumn{2}{c}{TREC-DL 21} 
& \multicolumn{2}{c}{TREC-DL 22} 
& \multicolumn{2}{c}{TREC-DL 23} \\
\cmidrule(lr){2-3} \cmidrule(lr){4-5}  \cmidrule(lr){6-7} \cmidrule(lr){8-9}  \cmidrule(lr){10-11} 
& \multicolumn{1}{c}{MAP@100} & \multicolumn{1}{c}{MRR@10} 
& \multicolumn{1}{c}{MAP@100} & \multicolumn{1}{c}{MRR@10} 
& \multicolumn{1}{c}{MAP@100} & \multicolumn{1}{c}{MRR@10} 
& \multicolumn{1}{c}{MAP@100} & \multicolumn{1}{c}{MRR@10} 
& \multicolumn{1}{c}{MAP@100} & \multicolumn{1}{c}{MRR@10} \\
\midrule
UMBRELA & \textbf{0.946} & 0.832 & \textbf{0.930} & \underline{0.837} & \textbf{0.921} & \textbf{0.868} & 0.905 & \textbf{0.774} & 0.849 & \textbf{0.844} \\
\midrule
monoT5 base & 0.883 & \underline{0.839} & 0.799 & 0.807 & 0.858 & 0.731 & 0.810 & 0.606 & 0.792 & 0.479 \\
monoT5 large & 0.910 & \textbf{0.841} & 0.825 & 0.702 & 0.863 & 0.786 & 0.825 & 0.625 & 0.825 & 0.741 \\
monoT5 3B & 0.895 & 0.832 & 0.868 & 0.692 & \underline{0.906} & 0.828 & 0.879 & 0.696 & 0.829 & 0.669 \\
Rank1-7B & 0.901 & 0.835 & 0.883 & 0.821 & 0.892 & 0.839 & 0.922 & 0.741 & \underline{0.889} & 0.817 \\
Rank1-14B & 0.892 & 0.763 & 0.887 & \textbf{0.847} & 0.890 & 0.829 & \textbf{0.937} & 0.747 & \textbf{0.909} & 0.797 \\
Rank1-32B & \underline{0.916} & 0.808 & \underline{0.889} & 0.834 & 0.886 & \underline{0.836} & \underline{0.927} & \underline{0.759} & \underline{0.899} & \underline{0.834} \\
\bottomrule
\end{tabular}
\end{table*}

\subsubsection{Datasets.}
Following recent studies~\cite{schnabel2025multi,upadhyay2024umbrela}, we use the TREC 2019--2023 Deep Learning (TREC-DL) tracks~\cite{craswell2019,craswell2020,2021craswell,craswell2022,craswell2023}; see statistics in Table~\ref{tab:data}. 
TREC-DL 2019--2020 and 2021--2023 are based on the MS MARCO V1 and V2 passage ranking collections, containing 8.8M and 138M passages, respectively. 
Following~\citep{boytsov2025positionalbiaslongdocumentranking}, we treat TREC-DL 2019--2020 as in-corpus evaluation, as all re-rankers used in this paper are trained on MS MARCO v1; in contrast, TREC-DL 2021--2023 can be considered an out-of-training-corpus evaluation, since TREC-DL 2021 onwards use the MS MARCO v2 collection.
Each dataset has four relevance levels: perfectly relevant (3), highly relevant (2), related (1), and irrelevant (0). 
We follow the official TREC-DL papers~\cite{craswell2019,craswell2020,2021craswell,craswell2022,craswell2023}, where passage relevance judgments are binarised by treating relevance levels $\geq$2 as positive.
All relevance labels in TREC-DL 2019–2021 were manually annotated by the U.S. National Institute of Standards and Technology (NIST) assessors, whereas TREC-DL 2022 and 2023 also include automatically propagated labels assigned to near-duplicate passages. 
Following~\cite{upadhyay2024umbrela}, we discard these propagated labels and retain only human annotations.

\subsubsection{Evaluation metric.}
We evaluate from two perspectives: judgment agreement and system ranking.
For agreement, following prior work~\cite{schnabel2025multi,meng2025query,faggioli2023perspectives}, we compute Cohen’s~$\kappa$ between human-annotated relevance judgments (qrels) and those predicted by re-ranker-based judges.
For system ranking, following~\cite{upadhyay2025large,upadhyay2024umbrela}, we compute Kendall’s~$\tau$ correlation between system orderings in terms of an IR evaluation metric derived from TREC assessors’ judgments and from each re-ranker-based judge; a system ranking on a TREC-DL dataset refers to the ordering of all submitted runs for that dataset.
For IR evaluation metrics, we follow prior studies and use MAP@100~\cite{dewan2025true,faggioli2023perspectives}, MRR@10~\cite{dewan2025true}.\footnote{We also include nDCG@10, with similar conclusions; due to space limitations, these results are available at \url{https://github.com/ChuanMeng/reranker-as-judge}.}


\subsubsection{Reference LLM-based judge.}
To provide a comparison point for re-ranker-based relevance judges, we use UMBRELA~\cite{upadhyay2024umbrela}, a state-of-the-art LLM-based relevance judge, as a reference.
Note that the goal of this paper is not to develop a state-of-the-art judge that outperforms UMBRELA.
Given a query and a document, UMBRELA prompts GPT-4o with a zero-shot descriptive, narrative, and aspects (DNA) prompting scheme~\cite{thomas2024large}.

\subsubsection{Implementation details.}
\label{sec:details}
For the re-ranker checkpoints, we use publicly available Hugging Face checkpoints for
monoT5-base (220M), monoT5-large (770M), monoT5-3B (3B),
RankLLaMA-7B, RankLLaMA-13B,
Rank1-7B, Rank1-14B, and Rank1-32B.\footnote{
All checkpoints are publicly available on Hugging Face.
Castorini (\url{https://huggingface.co/castorini}):
monot5-base-msmarco, monot5-large-msmarco, monot5-3b-msmarco,
rankllama-v1-7b-lora-passage, rankllama-v1-13b-lora-passage;
JHU-CLSP (\url{https://huggingface.co/jhu-clsp}):
rank1-7b, rank1-14b, rank1-32b.
}
%
For direct generation (RQ1), we record the special tokens generated by monoT5 and Rank1, following the default decoding settings from the original implementations.
For score thresholding (RQ2), for each re-ranker-based judge, we select the threshold that yields the best performance on one TREC-DL dataset and apply it to another as follows: 
TREC-DL 19$\rightarrow$20, TREC-DL 20$\rightarrow$19, TREC-DL 21$\rightarrow$22, TREC-DL 22$\rightarrow$21, and TREC-DL 22$\rightarrow$23.
For bias analysis (RQ3), we use Pyserini BM25~\citep{robertson1995okapi} with the default parameters k1=0.9, b=0.4.\footnote{\url{https://github.com/castorini/pyserini}}
BM25 retrieves the top 1,000 documents, which are then re-ranked by each re-ranker.
For UMBRELA~\citep{upadhyay2024umbrela}, we use graded relevance judgments generated by the original authors and binarise them; these judgments were kindly provided upon request.
All experiments are conducted on RTX 6000 Ada (48GB) GPUs.

\section{Results and Discussions}
\label{sec:result}

\subsection{Re-rankers as judges via direct generation}

To answer \ref{RQ1}, we evaluate re-ranker-based relevance judges adapted from monoT5~\cite{nogueira2020document} and Rank1~\cite{weller2025rank} via direct generation; we also use UMBRELA~\cite{upadhyay2024umbrela}, a state-of-the-art LLM-based relevance judge, as a reference. 
Note that direct generation is not applicable to RankLLaMA~\cite{ma2024fine}, which produces continuous relevance scores without producing special tokens (e.g., ``true'' or ``false''). 
Results are shown in Tables~\ref{tab:dg_cohen} (relevance judgment agreement in terms of Cohen’s~$\kappa$) and~\ref{tab:dg_kendall} (system ranking in terms of MAP@100 and MRR@10).
We have three main observations.

First, monoT5- and Rank1-based judges achieve comparable or higher agreement~(Cohen’s~$\kappa$) with TREC assessors than UMBRELA in roughly 40\% of the cases, as well as comparable or better system-ranking correlations than UMBRELA in approximately 50\% of the cases.
For instance, as shown in Table~\ref{tab:dg_cohen}, Rank1-7B achieves the same $\kappa$ value (0.450) as UMBRELA on TREC-DL 20 and a higher value (0.425) on TREC-DL 22. 
Similarly, as shown in Table~\ref{tab:dg_kendall}, monoT5-large achieves the highest Kendall’s $\tau$ correlation in terms of MRR@10 on TREC-DL 19 (0.841), while Rank1-14B attains the highest correlation in terms of MRR@10 on TREC-DL 20 (0.847) and MAP@100 on TREC-DL 22 (0.937); on TREC-DL 23, Rank1-14B again yields the best MAP@100 correlation (0.909).

%

Second, Rank1-based judges generally outperform monoT5-based ones and exhibit greater robustness across datasets.
In particular, monoT5 exhibits a clear performance drop on TREC-DL 22 and 23, whereas Rank1 remains relatively stable. 
For example, most Cohen’s~$\kappa$ values of monoT5 fall below 0.3 on TREC-DL 22 and 23, while Rank1-7B achieves the highest agreement on TREC-DL 22 (0.425) and Rank1-14B achieves the second-highest agreement on TREC-DL 23 (0.394).
Note that all re-rankers are trained on the MS MARCO v1 passage ranking corpus~\cite{bajaj2018ms}.
While TREC-DL 19--20 are based on MS MARCO v1, TREC-DL 21--23 adopt the MS MARCO v2 passage ranking corpus.
These observations suggest that reasoning-based Rank1 exhibits stronger robustness under corpus changes.

Third, scaling up re-rankers shows mixed results: it generally improves system-ranking performance, but does not consistently improve relevance judgment agreement.
For example, Rank1-14B and Rank1-32B achieve higher Kendall’s~$\tau$ correlations than Rank1-7B in most cases; however, scaling from 14B to 32B does not yield a clear advantage.
By contrast, Rank1-7B shows higher relevance agreement than Rank1-14B and Rank1-32B in 60\% of the cases.

\begin{table*}[th!]
\centering
\caption{
Relevance judgment agreement (Cohen’s $\kappa$) between TREC assessors and each re-ranker-based relevance judge adapted via \textit{score thresholding}.
The results of UMBRELA, state-of-the-art relevance judge, are included for reference.
The best value in each column is \textbf{boldfaced}, while the second best is \underline{underlined}.
}
\label{tab:st_cohen}
\begin{tabular}{l ccccc}
\toprule
\multirow{1}{*}{Method} & TREC-DL 19 & TREC-DL 20 & TREC-DL 21 & TREC-DL 22 & TREC-DL 23 \\
\midrule 
UMBRELA & \textbf{0.499} & \underline{0.450} & \textbf{0.492} & 0.421 & \textbf{0.418} \\
\midrule
monoT5 base & 0.350 & 0.307 & 0.341 & 0.240 & 0.269 \\
monoT5 large & 0.398 & 0.311 & 0.373 & 0.256 & 0.295 \\
monoT5 3B & 0.391 & 0.364 & 0.273 & 0.273 & 0.336 \\
RankLLaMA-7B & 0.400 & 0.371 & 0.387 & 0.232 & 0.311 \\
RankLLaMA-13B & 0.407 & 0.340 & 0.360 & 0.260 & 0.314 \\
Rank1-7B & 0.443 & \textbf{0.455} & \underline{0.482} & \textbf{0.430} & 0.386 \\
Rank1-14B & \underline{0.463} & 0.438 & 0.472 & 0.425 & \underline{0.406} \\
Rank1-32B & 0.454 & 0.417 & 0.461 & \underline{0.426} & 0.392 \\
\bottomrule
\end{tabular}
\end{table*}

\begin{table*}[th!]
\centering
\caption{
Kendall’s $\tau$ correlation coefficients between the system orderings induced by relevance judgments from TREC assessors and those predicted by each re-ranker-based relevance judge adapted via \textit{score thresholding}. 
Results are shown for system orderings based on MAP@100 and MRR@10.
The results of UMBRELA, state-of-the-art relevance judge, are included for reference.
The best value in each column is \textbf{boldfaced}, while the second best is \underline{underlined}.
}
\label{tab:st_kendall}
\setlength{\tabcolsep}{3.4pt} 
\begin{tabular}{l ccccccccccc}
\toprule
\multirow{2}{*}{Method} 
& \multicolumn{2}{c}{TREC-DL 19} 
& \multicolumn{2}{c}{TREC-DL 20} 
& \multicolumn{2}{c}{TREC-DL 21} 
& \multicolumn{2}{c}{TREC-DL 22} 
& \multicolumn{2}{c}{TREC-DL 23} \\
\cmidrule(lr){2-3} \cmidrule(lr){4-5} \cmidrule(lr){6-7} \cmidrule(lr){8-9} \cmidrule(lr){10-11}
& MAP@100 & MRR@10 & MAP@100 & MRR@10 & MAP@100 & MRR@10 & MAP@100 & MRR@10 & MAP@100 & MRR@10 \\
\midrule
UMBRELA & \textbf{0.946} & 0.832 & \textbf{0.930} & \textbf{0.837} & \textbf{0.921} & \textbf{0.868} & 0.905 & \textbf{0.774} & 0.849 & \textbf{0.844} \\
\midrule
monoT5 base & 0.889 & \textbf{0.846} & 0.795 & 0.668 & 0.872 & 0.784 & 0.776 & 0.570 & 0.808 & 0.487 \\
monoT5 large & 0.898 & 0.749 & 0.832 & 0.714 & 0.870 & 0.814 & 0.833 & 0.668 & 0.815 & 0.669 \\
monoT5 3B & 0.892 & 0.818 & 0.868 & 0.752 & \underline{0.902} & 0.842 & 0.887 & 0.727 & 0.835 & 0.719 \\
RankLLaMA-7B & 0.889 & 0.806 & 0.877 & 0.664 & 0.860 & 0.799 & 0.834 & 0.660 & 0.788 & 0.655 \\
RankLLaMA-13B & 0.892 & 0.781 & 0.876 & 0.651 & 0.856 & \underline{0.836} & 0.841 & \underline{0.768} & 0.785 & 0.700 \\
Rank1-7B & \underline{0.910} & \underline{0.835} & 0.884 & 0.795 & 0.881 & 0.844 & 0.919 & 0.725 & 0.886 & \underline{0.804} \\
Rank1-14B & 0.898 & 0.763 & 0.887 & 0.820 & 0.889 & 0.822 & \underline{0.924} & 0.746 & \textbf{0.906} & 0.797 \\
Rank1-32B & \underline{0.910} & 0.808 & \underline{0.890} & \underline{0.830} & 0.882 & 0.835 & \textbf{0.925} & 0.751 & \underline{0.896} & \textbf{0.844} \\
\bottomrule
\end{tabular}
\end{table*}

\begin{figure*}[th!]
  \centering
  \includegraphics[width=1\linewidth]{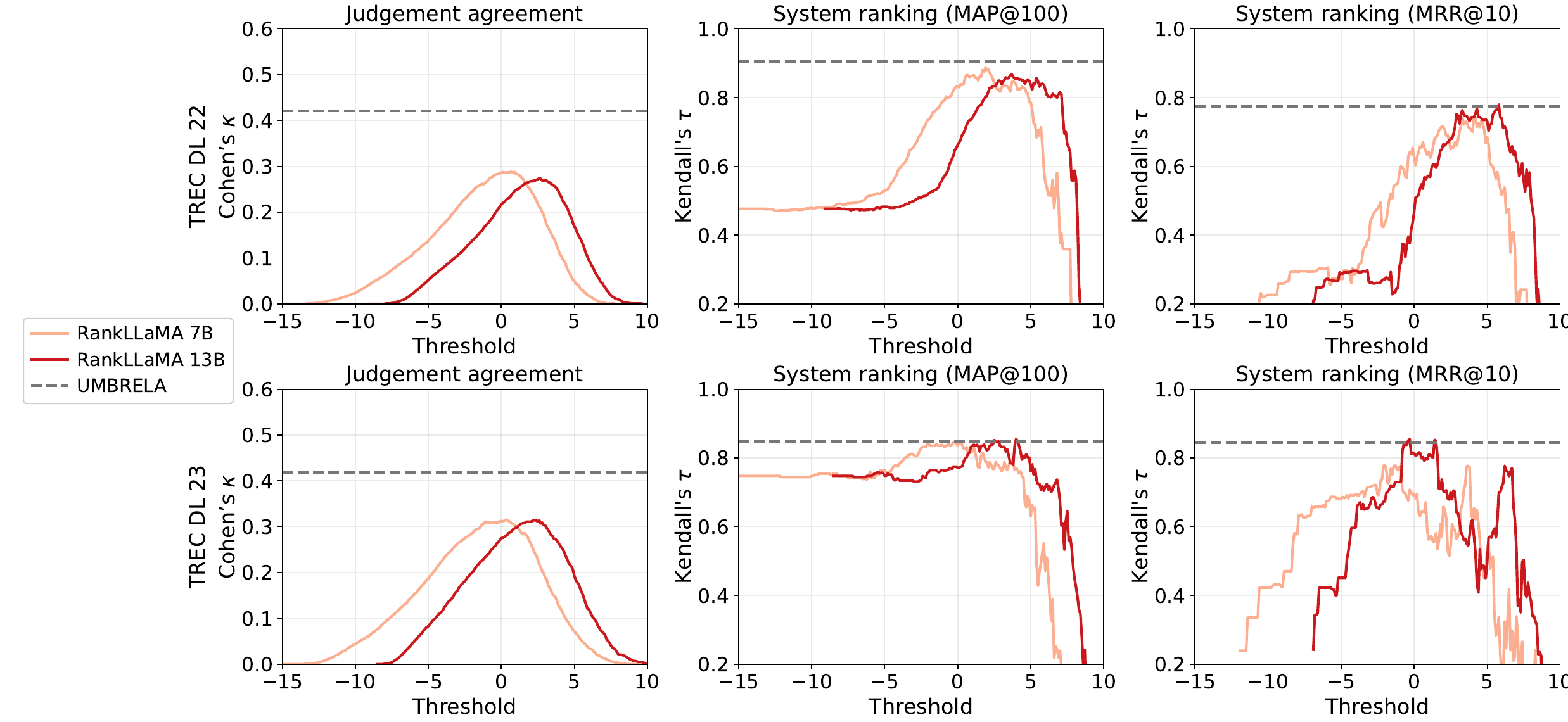}
  \caption{RankLLaMA's performance across thresholds on TREC-DL 22 and 23.
  The best-performing threshold from TREC-DL 22 is used for TREC-DL 23.
  }
    \label{fig:rankllama_threshold}
\end{figure*}

\begin{figure*}[th!]
  \centering
  \includegraphics[width=1\linewidth]{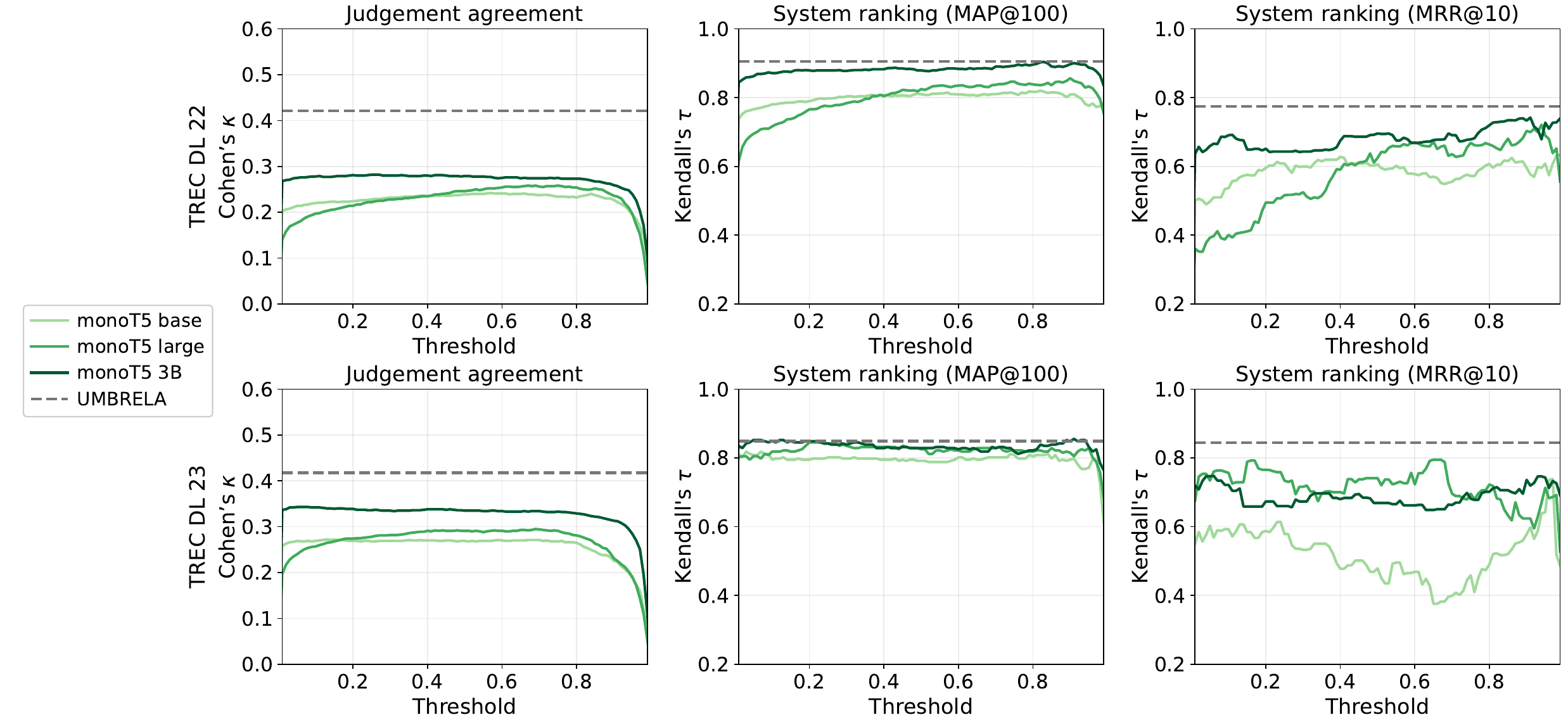}
  \caption{monoT5's performance across thresholds on TREC-DL 22 and 23.}
    \label{fig:monot5_threshold}
\end{figure*}

\begin{figure*}[th!]
  \centering
  \includegraphics[width=1\linewidth]{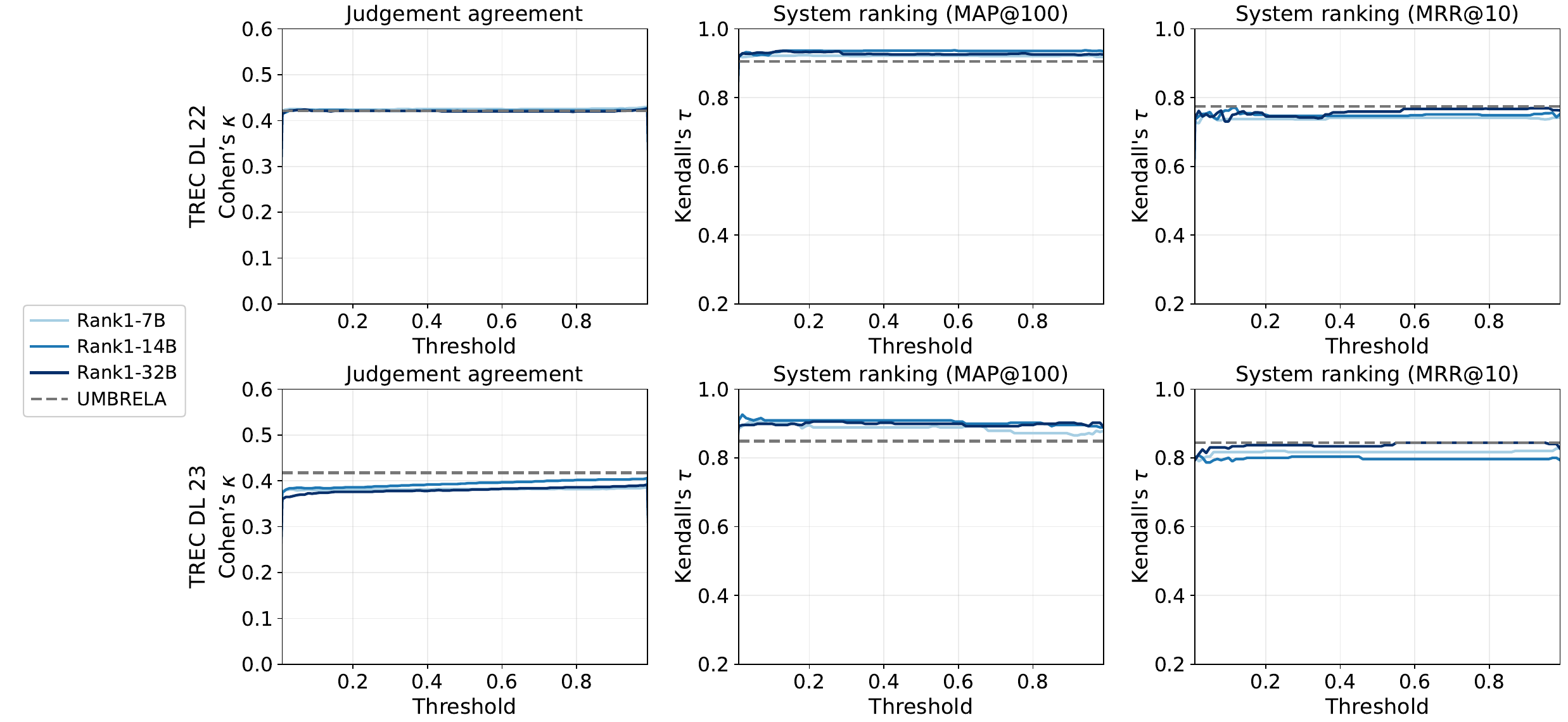}
  \caption{Rank1’s performance across thresholds on TREC-DL 22 and 23.}
    \label{fig:rank1_threshold}
\end{figure*}

\subsection{Re-rankers as judges via score thresholding}

To answer \ref{RQ2}, we evaluate re-ranker-based relevance judges adapted from monoT5~\cite{nogueira2020document}, RankLLaMA~\cite{ma2024fine}, and Rank1~\cite{weller2025rank} via score thresholding; we also use UMBRELA~\cite{upadhyay2024umbrela}, a state-of-the-art LLM-based relevance judge, as a reference. 
See Section~\ref{sec:details} for threshold selection details.
We report results on Tables~\ref{tab:st_cohen} (relevance judgment agreement in terms of Cohen’s~$\kappa$) and \ref{tab:st_kendall} (system ranking in terms of MAP@100 and MRR@10).

Similar to the results for \ref{RQ1}, re-ranker-based judges achieve higher performance than UMBRELA in roughly 40\% of the cases, in terms of both relevance judgment agreement and system-ranking correlations.
Rank1-based judges continue to demonstrate stronger and more robust performance than monoT5-based judges.
Scaling re-rankers again yields mixed results, tending to benefit system-ranking performance more than relevance judgment agreement.
Unlike the findings for \ref{RQ1}, scaling Rank1 to 32B provides more gains than the 14B and 7B variants under score thresholding.

\begin{table*}[!ht]
\centering
\caption{
Ranking performance of BM25 and BM25 with 8 re-rankers from 3 re-ranker families: monoT5 (220M, 770M, and 3B), RankLLaMA (7B and 13B), and Rank1 (7B, 14B, and 32B).
BM25 retrieves the top 1,000 items, which are subsequently re-ranked by each re-ranker.
Results are reported in terms of MRR@10 and MAP@100.
We also report Judge@10~\citep{pradeep2023rankzephyr}, which measures the fraction of items judged by human annotators among the top 10 ranked documents.
}
\label{tab:ranking_performance}
\setlength{\tabcolsep}{4pt} 
\begin{tabular}{l ccc ccc ccc}
\toprule
\multirow{2}{*}{Method} 
& \multicolumn{3}{c}{TREC-DL 19} 
& \multicolumn{3}{c}{TREC-DL 20} 
& \multicolumn{3}{c}{TREC-DL 21} \\
\cmidrule(lr){2-4}
\cmidrule(lr){5-7}
\cmidrule(lr){8-10}
& MAP@100 & MRR@10 & Judge@10
& MAP@100 & MRR@10 & Judge@10
& MAP@100 & MRR@10 & Judge@10 \\
\midrule
BM25 & 0.248 & 0.702 & 1.000 & 0.269 & 0.653 & 0.994 & 0.136 & 0.498 & 1.000  \\
\midrule
BM25+monoT5 base & 0.430 & 0.871  & 0.949 & 0.478 & 0.880 & 0.987 & 0.309 & 0.824 & 0.955  \\
BM25+monoT5 large & 0.439 & 0.873 & 0.949 & 0.479 & 0.831 & 0.976  & 0.313 & 0.833  & 0.955 \\
BM25+monoT5 3B & 0.437 & 0.867 & 0.965 &  0.522 & 0.886  & 0.989 & 0.306 & 0.793 & 0.957  \\
BM25+RankLLaMA-7B & 0.457 & 0.841 & 0.958  & 0.521 & 0.906 & 0.985 & 0.339 & 0.825 & 0.972 \\
BM25+RankLLaMA-13B & 0.451 & 0.895  & 0.977 & 0.512 & 0.861  & 0.978 & 0.342 & 0.851 & 0.972 \\
BM25+Rank1-7B & 0.402  &  0.810 & 0.847  & 0.450  & 0.790  & 0.824 & 0.272 & 0.778 & 0.757 \\
BM25+Rank1-14B & 0.399 & 0.754 & 0.809  & 0.453 & 0.736  & 0.856 & 0.263 & 0.735 & 0.713  \\
BM25+Rank1-32B & 0.416  & 0.828 & 0.830 & 0.467 & 0.774  & 0.880 & 0.289 & 0.750 & 0.777 \\
\bottomrule
\end{tabular}
\end{table*}

We make an additional observation: RankLLaMA-based judges generally underperform Rank1-based judges and do not show a clear advantage over monoT5-based ones.
To further investigate this behaviour, we analyse the performance of RankLLaMA, monoT5, and Rank1 across different thresholds on TREC-DL 22 and 23, as shown in Figures~\ref{fig:rankllama_threshold}, \ref{fig:monot5_threshold}, and \ref{fig:rank1_threshold}, respectively.
Note that the best-performing threshold from TREC-DL 22 is applied to TREC-DL 23~(see Section~\ref{sec:details} for details).
Two key findings emerge.
First, RankLLaMA exhibits substantial sensitivity to threshold selection for system-ranking performance, particularly for MRR@10.
For example, as shown in Figure~\ref{fig:rankllama_threshold}, the optimal threshold for RankLLaMA-13B in terms of MRR@10 is around 5 on TREC-DL 22, but shifts to approximately 0 on TREC-DL 23.
A similar degree of threshold sensitivity is also observed for monoT5 in terms of MRR@10.

Second, as shown in Figure~\ref{fig:rank1_threshold}, Rank1 demonstrates markedly stronger robustness to threshold variation than both RankLLaMA and monoT5.
We perform a closer inspection: Rank1’s predicted relevance scores (in the range $[0,1]$) are highly concentrated near the extremes: scores are often close to 1 (e.g., 0.99) or near 0.
In contrast, monoT5 produces a larger proportion of relevance scores in the middle of the $[0,1]$ range.
As a result, Rank1-based judges are substantially less sensitive to the choice of threshold.
The results suggest that highly concentrated relevance scores generalise more robustly across datasets under score thresholding.




\begin{figure*}[!ht]
  \centering
  \includegraphics[width=\linewidth]{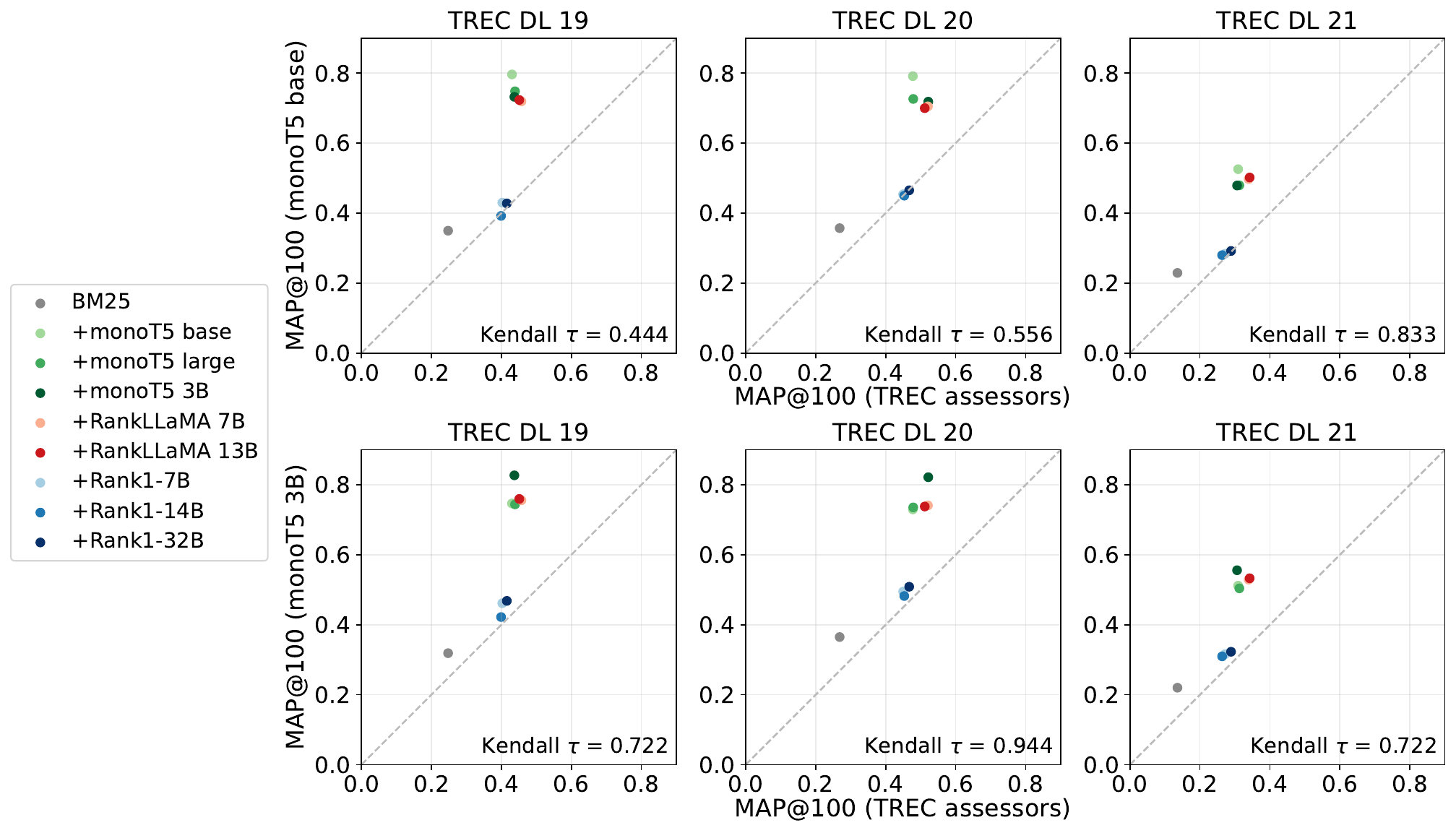}
  \caption{Scatter plots comparing system rankings based on relevance judgments from TREC assessors and monoT5 judges.}
    \label{fig:bias_monot5}
\end{figure*}

\begin{figure*}[!h]
  \centering
  \includegraphics[width=\linewidth]{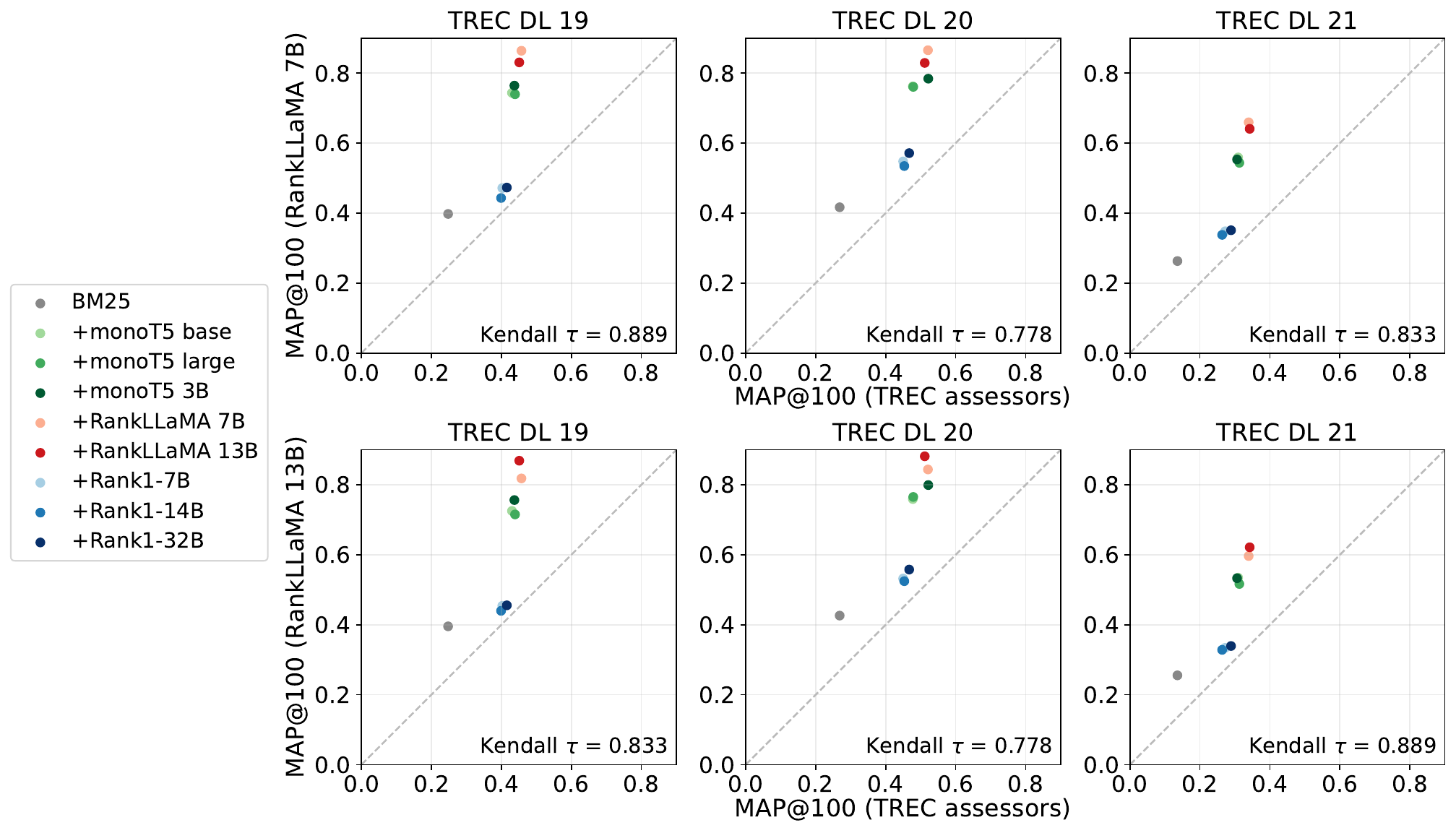}
  \caption{Scatter plots comparing system rankings based on relevance judgments from TREC assessors and RankLLaMA judges.}
  \label{fig:bias_rankllama}
\end{figure*}

\begin{figure*}[!ht]
  \centering
  \includegraphics[width=\linewidth]{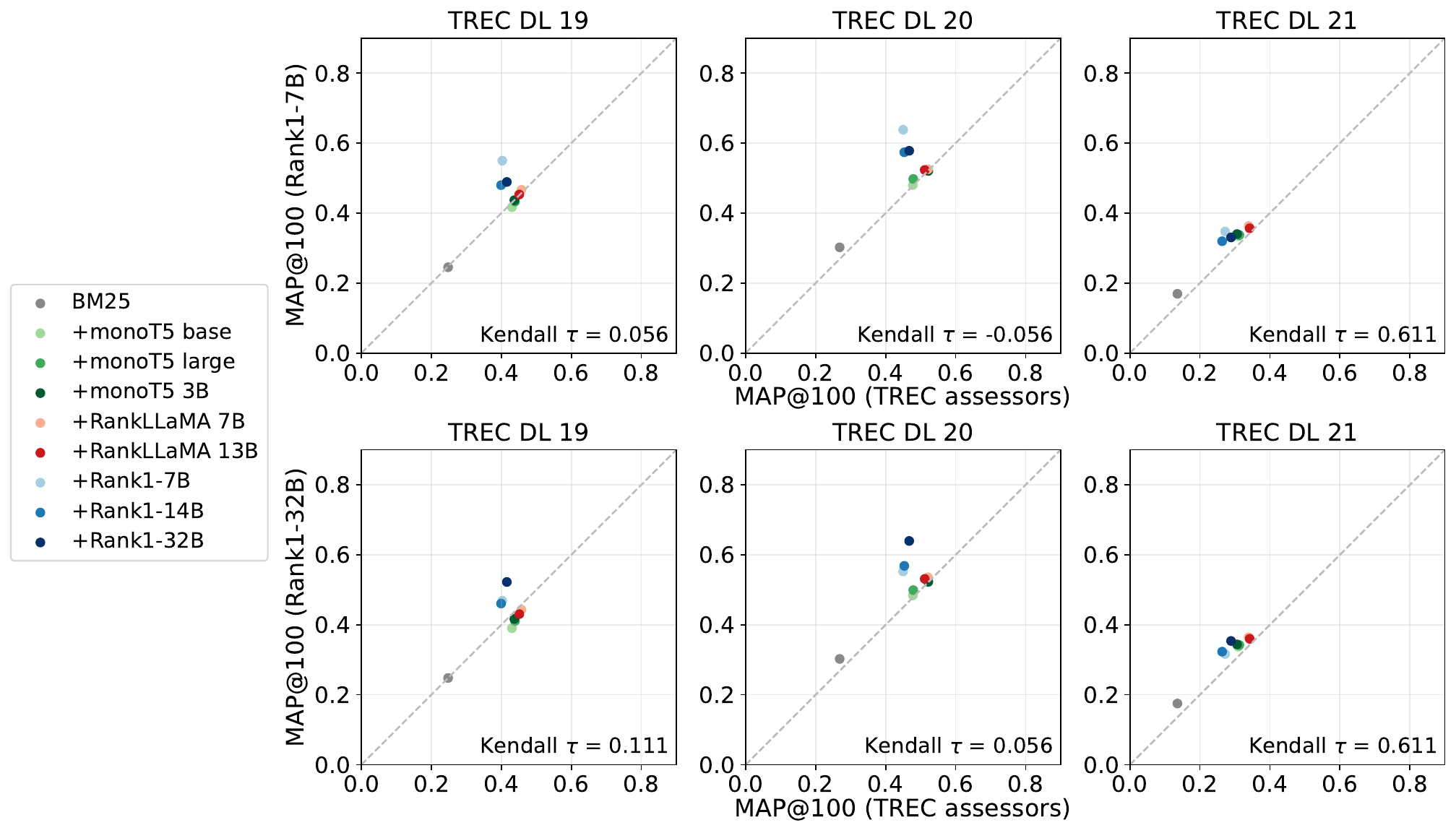}
  \caption{Scatter plots comparing system rankings based on relevance judgments from TREC assessors and Rank1-based judges.}
    \label{fig:bias_rank1}
\end{figure*}

\begin{figure*}[!ht]
  \centering
  \includegraphics[width=\linewidth]{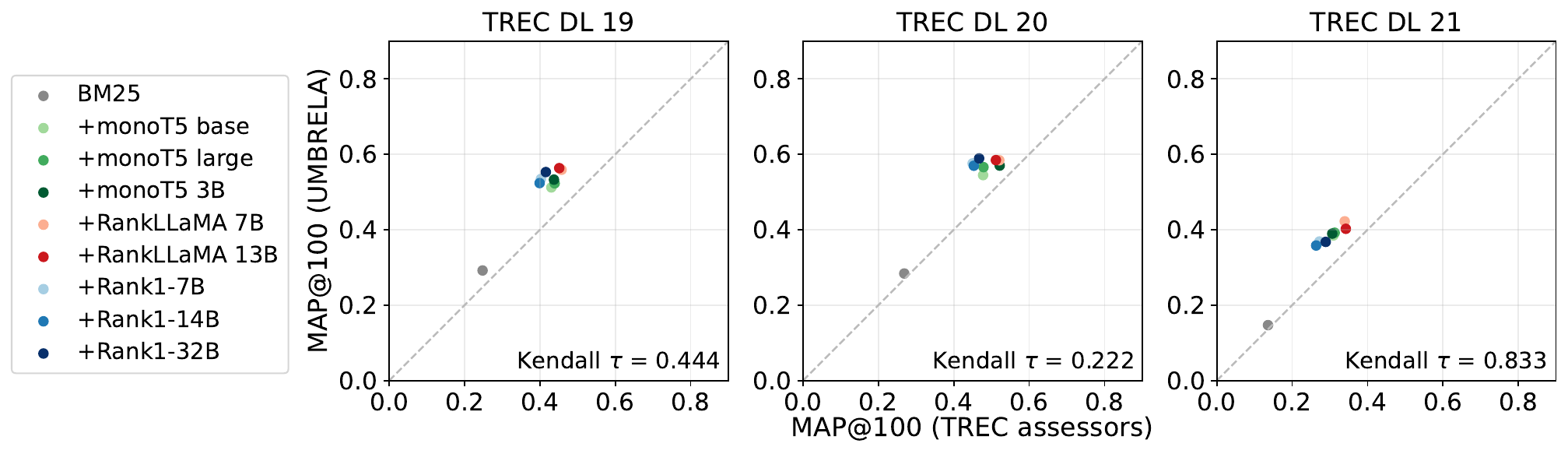}
  \caption{Scatter plots comparing system rankings based on relevance judgments from TREC assessors and UMBRELA.}
    \label{fig:bias_umbrella}
\end{figure*}

\subsection{Bias of re-ranker-based judges}

To address \ref{RQ3}, we curate 9 run files for each TREC-DL dataset: BM25 and BM25 followed by 8 each of re-rankers: monoT5 (220M, 770M, 3B), RankLLaMA (7B, 13B), and Rank1 (7B, 14B, 32B).
See Sections~\ref{sec-bias-method} and \ref{sec:details} for details.
Each re-ranker-based judge is used to evaluate and rank all systems, including both the re-ranked runs and the BM25 run, in terms of MAP@100.

As a sanity check, we first report the ranking performance of the 9 runs evaluated using TREC assessor–annotated relevance judgments on TREC-DL 19, 20, and 21 in Table~\ref{tab:ranking_performance}.
We observe that although Rank1 is shown to be an effective judge in \ref{RQ1}/\ref{RQ2}, it shows limited re-ranking quality under TREC assessor–annotated qrels.
E.g., BM25--Rank1-7B achieves MAP@100 scores of 0.402, 0.450, and 0.272 on TREC-DL 19, 20, and 21, respectively.
Its judge@10 values (0.847, 0.824, 0.757) are clearly lower than those of BM25--RankLLaMA-7B (0.958, 0.985, 0.972) on the same datasets.
This is likely because many relevant documents returned by Rank1 are not judged in the TREC qrels, leading to underestimated performance.
Our finding is consistent with \cite{weller2025rank}.

We then visualise the results using scatter plots comparing system rankings derived from relevance judgments of TREC assessors and those predicted by the re-ranker-based judges.
Due to space limitations, we show results on TREC-DL 19, 20 and 21, with judges based on monoT5-base/3B, RankLLaMA-7B/13B, and Rank1-7B/32B in Figures~\ref{fig:bias_monot5}, \ref{fig:bias_rankllama}, and~\ref{fig:bias_rank1}, respectively.
Note that monoT5 and Rank1 use direct generation (\ref{RQ1}), while RankLLaMA uses the thresholds found in \ref{RQ2}.
UMBRELA is also shown as a reference judge in Figure~\ref{fig:bias_umbrella}.
We make three main observations.

First, all re-ranker-based judges exhibit clear self-preference: each judge tends to assign the highest scores to its own re-ranker and to other re-rankers within the same family.
E.g., in Figure~\ref{fig:bias_rankllama}, the RankLLaMA-7B judge ranks RankLLaMA-7B first and RankLLaMA-13B second, while the RankLLaMA-13B judge shows the reverse pattern. 

Second, both monoT5 and RankLLaMA judges tend to rate each other as the second-best family after themselves.
They consistently rate Rank1 much lower, in some cases placing its re-ranking performance close to that of BM25 alone. This indicates a cross-family bias: non-reasoning-based re-rankers favour other non-reasoning-based families over the reasoning-based Rank1 family. 

Third, UMBRELA and Rank1 accurately predict BM25’s performance (BM25’s grey points lie close to the diagonal), whereas monoT5- and RankLLaMA-based judges tend to overestimate it. 



\vspace*{-2.5mm}
\section{Related Work}
Recent studies on using LLMs as relevance judges can be broadly grouped into three lines of research.
First, a large body of work focuses on methodologies for predicting relevance judgments, including prompt design~\cite{farzi2025criteria,thomas2024large,faggioli2023perspectives,dewan2025llm}, fine-tuning LLMs on human-labelled data~\cite{gienapp2025topic,meng2025query,abbasiantaeb2025improving}, in-context example selection~\cite{mckechnie2025context}, pipeline design~\citep{schnabel2025multi}, document summarisation~\citep{mohtadi2025effect}, and ensembling multiple LLM-based judges~\cite{rahmani2025judgeblender}.
Second, many studies analyse the behaviour of LLM-based judges, like investigating cognitive biases~\citep{chen2024ai}, preservation of ranking differences amongst top systems~\cite{otero2025limitations}, and sensitivity to input variations~\cite{arabzadeh2025human,alaofi2024llms}.
Third, several studies explore practical applications, including completing missing relevance judgments~\cite{abbasiantaeb2025improving,abbasiantaeb2024can,upadhyay2024llms,gienapp2025topic}, constructing fully synthetic test collections~\cite{rahmani2025syndl,rahmani2024synthetic}, and combining manual and LLM-generated annotations to construct test collections under limited budgets~\cite{takehi2025llm}.

The most closely related work includes \citep{soboroff2025don,clarke2024llm}, \citep{frobe2025large}, and \citep{macavaney2023one}.
\citep{soboroff2025don,clarke2024llm} discuss the conceptual connection between ranking and relevance judging; in contrast, we provide a systematic empirical study by reproducing re-rankers as relevance judges.
\citep{frobe2025large} analyse evaluation bias and show that LLM-based judges (built following UMBRELA~\citep{upadhyay2024umbrela}) favour LLM-based search systems; in contrast, our work focuses on using re-rankers as relevance judges.
\citep{macavaney2023one} use a pairwise re-ranker in a one-shot setting with a known relevant document per query; in contrast, we study pointwise re-rankers in a more realistic setting without prior relevance information.


\vspace*{-2mm}
\section{Conclusions \& Future Work}
In this work, we have reproduced 8 re-rankers from three families of varying sizes as relevance judges.
To enable this, we have devised two adaptation strategies, direct generation and score thresholding, allowing re-rankers to function as relevance judges.
Moreover, we have analysed evaluation bias exhibited by re-ranker-based judges.
Overall, experiments on TREC-DL 2019--2023 show that most re-ranking findings carry over to the
``re-ranker-as-relevance-judge'' setting.
Specifically, we observe that:
\begin{enumerate*}[label=(\roman*)]
\item Under both strategies, re-ranker-based judges exhibit competitive performance, outperforming UMBRELA, a state-of-the-art LLM-based judge, in around 40\%--50\% of the cases.
Amongst the re-ranker-based judges, reasoning-based Rank1 generally performs best and shows greater robustness across datasets.
Scaling up re-ranker model size tends to be beneficial, but the gains are not always consistent. 
\item Score thresholding is particularly effective for re-rankers whose relevance scores are highly concentrated near the extremes; Rank1's predicted relevance scores are often very close to 1 or near 0, making it less sensitive to threshold selection than other re-rankers.
\item Re-ranker-based judges exhibit pronounced biases, favouring their own and same-family re-rankers, showing cross-family bias, and often overestimating BM25’s retrieval quality.
\end{enumerate*}

Our study further suggests that relevance judging is not a fundamentally new task but rather a specific case of \textit{relevance prediction}, consistent with prior arguments~\cite{soboroff2025don,clarke2024llm}.
This connection implies that our community can benefit from building relevance judges \textit{on top of} well-established relevance prediction methods (e.g., a re-ranker), rather than designing new relevance judges from scratch. 
By reusing existing models and data, we can reduce duplicated effort and avoid underutilisation of existing resources~\cite{jiang2025landscape}.

Future work includes:
\begin{enumerate*}[label=(\roman*)]
\item Mitigating biases in re-ranker-based judges; inspired by recent work~\cite{gienapp2025topic}, adding adapter parameters to adapt a re-ranker as a relevance judge appears to be a promising direction.
\item While many widely used IR metrics (e.g., Precision, Recall, MRR and MAP) operate under binary judgments, proposing adaptation strategies that support graded labels is a natural next step.
\item It would be valuable to adapt re-rankers beyond pointwise ones (e.g., listwise ones~\cite{yang2025rank,liu2025reasonrank,rathee2025guiding}), and use other datasets with reasoning-intensive queries (e.g., BRIGHT~\citep{hongjinbright}) or conversational queries~\citep{mo2025uniconv,mo2025convmix,meng2025bridging,mo2025conversational}.
\item Rank1’s potentially underestimated performance~(see Table~\ref{tab:ranking_performance}) highlights the need for our community to re-examine unjudged documents returned by new rankers.
\end{enumerate*}

\begin{acks}
This research was supported by a Turing AI Acceleration Fellowship funded by the Engineering and Physical Sciences Research Council (EPSRC), under grant number EP/V025708/1.
It was also supported by the Dutch Research Council (NWO), under project numbers 024.004.022, NWA.1389.20.\-183, and KICH3.LTP.20.006, and the European Union under grant agreements No. 101070212 (FINDHR) and No. 101201510 (UNITE).

All content represents the opinion of the authors, which is not necessarily shared or endorsed by their respective employers and/or sponsors.
\end{acks}


\bibliographystyle{ACM-Reference-Format}
\balance
\bibliography{references}

\end{document}